\documentclass[12pt]{article}
\usepackage{graphicx}

\def\ra {\rightarrow}
\def\qbq {\overline qq}
\def\qb {\overline q}
\def\ubu {\overline uu}
\def\dbd {\overline dd}

\def\kbk {\overline KK}
\def\pp {\pi\pi}
\def\qbqg {\overline qqg}
\def\sbs {\overline ss}
\def\journal{\topmargin 0.0in   \oddsidemargin 0in
        \headheight 0pt \headsep 0pt
        \textwidth 6.5in 
\textheight 9in 
        \marginparwidth 1.5in
        \parindent 2em
        \parskip .5ex plus .1ex         \jot = 1.5ex}
\journal

\def\ra{\rightarrow}
\begin{document}
\begin{titlepage}

\noindent July 26, 2005      \hfill    LBNL-57700\\
\noindent Rev. August 19, 2005

\begin{center}

\vskip .5in

{\large \bf Chiral Suppression of Scalar Glueball Decay}

\vskip .5in

Michael S. Chanowitz\footnote{Email: chanowitz@lbl.gov}

\vskip .2in

{\em Theoretical Physics Group
     Ernest Orlando Lawrence Berkeley National Laboratory\\
     University of California\\
     Berkeley, California 94720}
\end{center}

\vskip .25in

\begin{abstract}

Because glueballs are $SU(3)_{Flavor}$ singlets, they are expected to
couple equally to $u,d$, and $s$ quarks, so that equal coupling
strengths to $\pi^+\pi^-$ and $K^+K^-$ are predicted. However, we show
that chiral symmetry implies the scalar glueball amplitude for $G_0
\rightarrow \overline qq$ is proportional to the quark mass, so that
mixing with $\sbs$ mesons is enhanced and decays to $K^+K^-$ are
favored over $\pi^+\pi^-$.  Together with evidence from lattice
calculations and from experiment, this supports the hypothesis that
$f_0(1710)$ is the ground state scalar glueball.

\end{abstract}

\end{titlepage}

\newpage

\renewcommand{\thepage}{\arabic{page}}
\setcounter{page}{1}

{\it 1. Introduction. --- } The existence of gluonic states is a
quintessential prediction of Quantum Chromodynamics (QCD). The key
difference between Quantum Electrodynamics (QED) and QCD is that
gluons carry color charge while photons are electrically
neutral. Gluon pairs can then form color singlet hadronic bound
states, ``glueballs,'' like mesons and baryons, which are color
singlet bound states of valence quarks.\cite{kuti} Because of
formidable experimental and theoretical difficulties, it is
frustrating, though not surprising, that this simple, dramatic
prediction has resisted experimental verification for more than two
decades.  Quenched lattice simulations predict that the mass of the
lightest glueball, $G_0$, a scalar, is near $\simeq 1.65$
GeV,\cite{morningstar} but the prediction is complicated by mixing
with $\overline qq$ mesons that require more powerful computations.
Experimentally the outstanding difficulty is that glueballs are not
easily distinguished from ordinary $\overline qq$ mesons, themselves
imperfectly understood. This difficulty is also exacerbated if mixing
is appreciable.

The most robust identification criterion, necessary but not
sufficient, is that glueballs are extra states, beyond those of the
$\overline qq$ meson spectrum. This is difficult to apply in practice,
though ultimately essential. In addition, glueballs are expected to be
copiously produced in gluon rich channels such as radiative $J/\psi$
decay, and to have small two photon decay widths. These two
expectations are encapsulated in the quantitative measure
``stickiness,''\cite{stickiness} which characterizes the relative
strength of gluonic versus photonic couplings.  

Another popular criterion is based on the fact that glueballs are
$SU(3)_{\rm Flavor}$ singlets which should then couple equally to
different flavors of quarks.  However we show here that the amplitude
for the decay of the ground state scalar glueball to quark-antiquark
is proportional to the quark mass, ${\cal M}(G_0 \ra \qbq) \propto
m_q$, so that decays to $\overline ss$ pairs are greatly enhanced over
$\overline uu +\overline dd$, and mixing with $\sbs$ mesons is
enhanced relative to $\ubu + \dbd$. We exhibit the result at leading
order and show that it holds to all orders in standard QCD
perturbation theory.

The result has a simple nonperturbative physical explanation,
similar, though different in detail, to the well known enhancement of
$\pi \ra \mu \nu$ relative to $\pi \ra e \nu$. For $m_q=0$ chiral
symmetry requires the final $q$ and $\overline q$ to have equal
chirality, hence unequal helicity, so that in the $G_0$ rest frame
with $z$ axis in the quark direction of motion, the total $z$
component of spin is nonvanishing, $|S_Z| = 1$. Because the ground
state $G_0$ $gg$ wave function is isotropic ($L=S=0$), the $\qbq$
final state is pure s-wave,\footnote{Integrating over the gluon
direction to project out the s-wave $gg$ wave function is equivalent
to integrating over the final quark direction with the initial gluon
direction fixed.}  $L=0$.  The total angular momentum is zero, and
since there is no way to cancel the nonvanishing spin contribution,
the amplitude vanishes. With one power of $m_q\neq 0$, the $q$ and
$\overline q$ have unequal chirality and the amplitude is allowed.

The enhancement is substantial, since $m_s$ is an order of magnitude
larger than $m_u$ and $m_d$.\cite{mq} But for scalar glueballs of mass
$\simeq 1.5 - 2$ GeV, $\Gamma(G_0 \ra \sbs)$, is suppressed of order
$(m_s/m_{G_0})^2$, so that it may be smaller than the nominally higher
order $G_0 \ra \qbqg$ process, which {\em is} $SU(3)_{\rm Flavor}$
symmetric. We find that the soft and collinear quark-gluon 
singularities of $G_0 \ra \qbqg$ vanish for $m_q = 0$, as they must 
if $G_0 \ra \qbq$ is to vanish at one loop order for $m_q = 0$. 
Unsuppressed, flavor-symmetric $G_0 \ra \qbqg$ decays are dominated
by configurations in which the gluon is well separated from
the quarks, which hadronize predominantly to multi-body final
states. 
$G_0$ may also decay flavor symmetrically to multigluon final states
($n \geq 3$), via the three and four gluon components of $G^2$ in
equation (1), by higher dimension operators that arise
nonperturbatively, or by higher orders in perturbation theory. The IR
singularities of $G_0 \ra ggg$ are cancelled by virtual corrections to
the $G_0$ wave function, while configurations with three (or more)
well separated gluons hadronize to multihadron final states.
The enhancement of $\sbs$ relative to $\ubu + \dbd$ is then
most strongly reflected in two body decays: we
expect $K^+K^-$ to be enhanced relative to $\pi^+ \pi^-$, while
multibody decays are more nearly flavor symmetric.

Glueball decay to light quarks 
and/or gluons 
cannot be computed reliably in any
fixed order of perturbation theory. However, the predicted ratio,
$\Gamma(G_0 \ra \sbs) / \Gamma(G_0 \ra \ubu + \dbd) \gg 1$, is
credible since it follows from an analysis to all orders in
perturbation theory and, in addition, from a physical argument that
does not depend on perturbation theory. The implication that $
\Gamma(G_0 \ra K^+K^-) \gg \Gamma(G_0 \ra \pi^+ \pi^-)$ is less secure
and is best studied on the lattice.  Remarkably, it is supported by an
early quenched study of $G_0$ decay to pseudoscalar meson pairs for
two ``relatively heavy'' $SU(3)_{\rm Flavor}$ symmetric values of
$m_q$, corresponding to $m_{\rm PS} \simeq 400$ and $\simeq 630$
MeV.\cite{weingarten} Linear dependence on $m_q$ implies quadratic
dependence on $m_{\rm PS}$,\cite{gmor} which is consistent at
$1\sigma$ with the lattice computations.\cite{weingarten} Chiral
suppression could then be the physical basis for 
the unexpected and unexplained lattice result.  With subsequent
computational and theoretical advances in lattice QCD, it should be
possible today to verify the earlier study and to extend it to smaller
values of $m_{PS}$, nearer to the chiral limit and to the physical
pion mass.  If the explanation is indeed chiral suppression, then the
couplings of higher spin glueballs should be approximately flavor
symmetric and independent of $m_{\rm PS}$, a prediction which can also
be tested on the lattice.

Enhanced strange quark decay changes the expected experimental
signature and supports the hypothesis that $f_0(1710)$ is
predominantly the ground state scalar glueball. This identification
was advocated by Sexton, Vaccarino, and Weingarten,\cite{weingarten}
and is even more compelling today in view of recent results from
$J/\psi$ decay obtained by BES\cite{besa,besb} --- see \cite{beijing} 
for an overview of the experimental situation. 

In section 2 we compute ${\cal M}(G_0 \ra \qbq)$ at leading order for
massive quarks, with the expected linear dependence on $m_q$. In
section 3 we show that ${\cal M}(G_0 \ra \qbq) \propto m_q$ to any
order in $\alpha_S$.  In section 4 we describe the infrared
singularities of ${\cal M}(G_0 \ra \qbqg)$. We conclude with
a brief discussion, including experimental implications.

{\it 2. $G_0 \ra \qbq$ at leading order. --- }
Consider the decay of a scalar glueball $G_0$
with mass $M_G$ to a $\overline qq$ pair\footnote
{
Elastic scattering, $gg\ra gg$, contributes to the glueball wave
function, not to the decay amplitude. The dissociation process, $gg\ra
\qbq + \qbq$, in which each gluon makes a transistion to a color-octet
$\qbq$ pair, is kinematically forbidden for $m_q \neq 0$; an additional
gluon exchange is required to allow it to proceed on-mass-shell, which
is therefore of order $g^4$ in the amplitude.
}
with quark mass $m_q$.  
The effective glueball-gluon-gluon coupling is parameterized by 
$$
{\cal L}_{eff} = f_0G_0 G_{a\mu\nu}G_a^{\mu\nu}  \eqno{(1)}
$$ where $ G_0$ is an interpolating field for the glueball,
$G_{a\mu\nu}$ is the gluon field strength tensor with color index $a$,
and $f_0$ is an effective coupling constant with dimension $1/M$ that
depends on the $G_0$ wave function. The $gg \ra \overline qq$
scattering amplitude can be written as
$$
{\cal M}(gg \ra \overline qq) =
\epsilon_{1\mu}\epsilon_{2\nu} {\cal M}^{\mu\nu}(gg \ra \overline qq)
                                                \eqno{(2)}
$$ 
where $\epsilon_{i\mu} =\epsilon_{\mu}(p_i,\lambda_i) $ with $i=1,2$
are the polarization vectors for massless constituent gluons with four
momentum $p_i$ and polarization $\lambda_i$.  Using equations (1) and 
(2), summing over the polarizations $\lambda_i$, and averaging over the
gluon direction in the $G_0$ rest frame to project out the s-wave, we obtain
$$
{\cal M}(G_0 \ra \overline qq) = {f_0\over 4\pi}
              \int d\Omega X^{\mu\nu}{\cal M}_{\mu\nu}(gg \ra \overline qq),
                                     \eqno{(3)}
$$ 
where 
 $
X^{\mu\nu}= 2p_2^{\mu}p_1^{\nu} - M_G^2 g^{\mu\nu}
$
projects out the $| (++)\ +\ (--) >$ helicity state that couples 
to $G_{a\mu\nu}G_a^{\mu\nu}$ in equation (1). 

From the lowest order Feynman diagrams we obtain 
$$
X^{\mu\nu}{\cal M}_{\mu\nu} = -{32\pi\sqrt{2}\alpha_S \over 3}
         \ {m_q\over 1 - \beta^2 {\rm cos}^2\theta} 
          \ \overline u(p_3,h_3) v(p_4,h_4) \delta_{ij}  \eqno{(4)}
$$
where $u_3, v_4$ are the $q,\overline q$ spinors for quark and
antiquark with center of mass momenta $p_3,p_4$, helicities $h_3,h_4$,
color indices $i,j$ and center of mass velocity $\beta$. 
Equation (4) includes a color factor from the 
color singlet $gg$ wave function, 
$$
C_{ij}={\delta_{a,b}\over \sqrt{8}}{\lambda^a_{ik}\over 2}
          {\lambda^{b}_{kj}\over 2} 
	  = {\sqrt{2}\over 3}\delta_{ij}          \eqno{(5)}
$$

Performing the angular integration, the decay amplitude is
$$ {\cal M}(G_0 \ra
\overline qq) = -f_0\alpha_S\ {16\pi\sqrt{2} \over 3}
                      \ {m_q\over \beta} \ {\rm
log} {1 + \beta \over 1 - \beta} \ \overline u_3 v_4 \delta_{ij}.
\eqno{(6)}
$$ 
Squaring equation (6), summing
over quark helicities and color indices, and performing the phase
space integration, the decay width is
$$
\Gamma(G_0 \ra \overline qq) = {16 \pi\over 3} \ \alpha_S^2 f_0^2
      m_q^2 M_G\\ \beta\ {\rm log}^2 {1 + \beta \over 1 - \beta}.   
                                   \eqno{(7)}
$$ 

Notice that an explicit factor $m_q$ appears in the $gg \ra \qbq$
amplitude, equation (4), which is not averaged over the initial gluon
direction and which clearly has contributions from higher partial
waves, $J > 0$. It may then appear that chiral suppression applies not
just to spin 0 glueballs but also to glueballs of higher spin. However
when equation (4) is squared and the phase space integration is
performed, a factor $1/m_q^2$ results from the $t$ and $u$ channel
poles, which cancels the explicit factor $m_q^2$ in the numerator, so
that the total cross section $\sigma(gg\ra \qbq)$ does not vanish in
the chiral limit, because of the $J>0$ partial waves.

{\it 3. $G_0 \ra \qbq$ to all orders. --- }
We now show that ${\cal M}(G_0 \ra \overline qq)$ vanishes to all 
orders in perturbation theory for $m_q=0$.  
Consider the Lorentz invariant amplitude 
$$
{\cal M}_X(p_1,p_2,p_3,p_4)= X^{\mu\nu}{\cal M}_{\mu\nu}  \eqno{(8)}
$$ 
where ${\cal M}_{\mu\nu}$ is defined in 
equation (2) and $X^{\mu\nu}$ below (3). The perturbative expansion 
for ${\cal M}_X$ 
is a sum of terms arising from Feynman diagrams with 
arbitrary numbers of loops. After evaluation of the loop 
integrals, regularized as necesary, ${\cal M}_X$ is 
a sum of terms,
$$
{\cal M}_X= \sum_i {\overline u}(p_3,\chi_3) \Gamma_i u(p_4,\chi_4),   
                                       \eqno{(9)}
$$ 
where $u_3,u_4$ are respectively massless fermion and antifermion 
spinors\cite{ks} of chirality 
$\chi_3, \chi_4$. 
The $\Gamma_i$ are $4 \times 4$ matrices, each 
a product of $n_i$ momentum-contracted Dirac matrices, 
$$
\Gamma_i = \not k_{i1} \not k_{i2} \ldots \not k_{in_{i}}   \eqno{(10)}
$$
where each $k_{ia}$ is one of the external four-momenta, 
$p_1,p_2,p_3,p_4$. 

Chiral invariance for $m_q=0$ implies that 
the number of factors, $n_i$, in equation (10) is always odd. 
Since all external momenta vanish and the spinors 
obey $\not p_3 u_3= \not p_4 u_4= 0$, by suitably
anticommuting the $\not k_{ia}$, each term in equation (9) can be reduced
to a sum of terms linear in $\not p_1$ and $\not p_2$, which 
we choose to be symmetric and antisymmetric, 
$$
 {\overline u}(p_3,\chi_3) \Gamma_i u(p_4,\chi_4) = 
    {\overline u}(p_3,\chi_3)  [S_i(s,t,u) (\not p_1 +\not p_2)
  + A_i(s,t,u) (\not p_1 -\not p_2)] u(p_4,\chi_4).   \eqno{(11)}
$$
The coefficients $A_i,S_i$ are Lorentz invariant functions of the 
Mandelstam variables 
$s,t,u$.
Since $p_1 +p_2= p_3+p_4$, the symmetric term vanishes and 
equation (9) reduces to 
$$
{\cal M}_X= A(s,t,u)\  {\overline u}(p_3,\chi_3)
 (\not p_1 -\not p_2) u(p_4,\chi_4).   \eqno{(12)}
$$
where 
$
A(s,t,u) =  \sum_i A_i(s,t,u)
$.

Next consider the integration over the gluon direction, equation
(3). In the $G_0$ rest frame with the z-axis chosen 
along the quark direction of motion, $\hat z = \hat p_3$, we 
integrate over $d\Omega = d^2\hat p_1 =
d\phi_1 d{\rm cos}\theta_1$. The Mandelstam variables are then $s=M_G^2$
and $u,t = -{1\over 2}M_G^2 (1 \pm {\rm cos}\theta_1)$. Since the color
and helicity components of the $G_0$ wave function are symmetric 
under interchange of
the two gluons, Bose symmetry requires $A(s,t,u)$ to be odd under $p_1 
\leftrightarrow p_2$. In our chosen coordinate system $A$ is a function 
only of ${\rm cos}\theta_1$, 
and must therefore be odd, 
$
A(-{\rm cos}\theta_1) = -A({\rm cos}\theta_1)
$. 
But evaluating $\overline u_3(\not p_1 -\not p_2) u_4$ 
explicitly\cite{ks} we find 
$$
\overline u_3(\not p_1 -\not p_2) u_4 = M_G^2\ e^{-i\phi_1}\ {\rm sin}\theta_1
   \eqno{(13)}
$$ 
which is {\em even} in ${\rm cos}\theta_1$, while the azimuthal
factor, $e^{-i\phi_1}$, provides the required oddness under $p_1
\leftrightarrow p_2$: $e^{-i\phi_1} \ra e^{-i(\phi_1+\pi)} =
-e^{-i\phi_1}$.  Consequently the integral $\int d{\rm
cos}\theta_1{\cal A}$ vanishes, and ${\cal M}(G_0 \ra \overline qq)=0$
to all orders in the chiral limit.  In fact, because of our choice of
axis, $\hat z= \hat p_3$, the integral over $\phi_1$ also
vanishes. For other choices of $\hat z$ the azimuthal and polar integrals do
not vanish separately, but the full angular integral, $\int d^2 \hat
p_1$, vanishes in any case.

For nonvanishing quark mass, $m_q \neq 0$, chirality-flip amplitudes
contribute. With one factor of $m_q$ from the
fermion line connecting the external quark and antiquark, the
$\Gamma_i$ matrices in equation (9) include products of even
numbers of Dirac matrices, i.e., $n_i$ in equation (10) may be 
even. Beginning in order $m_q$ there are then nonvanishing
contributions to ${\cal M}(G_0 \ra \overline qq)$, like the leading
order term shown explicitly in equation (6).

The vanishing azimuthal integration for $\hat z = \hat p_3$ reflects
the physical argument given in the introduction. The factor
$e^{-i\phi_1}$ corresponds to $S_Z=1$ from the aligned spins of the
$q$ and $\overline q$, while the absence of a compensating factor in
$\cal A$ is due to the projection of the orbital s-wave by the $\int
d^2 \hat p_1$ integration and the absence of
spin-polarization in the initial state.

{\it 4. Infrared singularities of $G_0 \ra \qbqg$. --- }
Although it is of order $\alpha_S^3$, $\Gamma(G_0 \ra \qbqg)$ is not
chirally suppressed and may therefore be larger than $\Gamma(G_0 \ra
\sbs)$, which is of order $\alpha_S^2\times m_s^2/m_G^2$.  Setting
$m_q=0$ we evaluated the 13 Feynman diagrams using the
helicity spinor method\cite{ks} with 
numerical evaluation of the 9 dimensional integral:
$$
\begin{array}{cclr}
\Gamma(G_0 \ra \qbqg )&=& \sum_{h_3,h_4,\lambda_5} 
          \int_{\rm PS}|{\cal M}(G_0 \ra \overline q_3q_4g_5)|^2 & \\
  &=& {f_0^2\over 16\pi^2}\sum_{h_3,h_4,\lambda_5} 
          \int_{\rm PS}\int d\Omega_1\int d\Omega_1^{\prime}
             \ \epsilon_{5\alpha}^*
            X_{\mu\nu}{\cal M}^{\mu\nu\alpha}(g_1g_2 \ra \overline q_4q_3g_5)
                & \\
       & &  \hspace{1.7in} \times \ \epsilon_{5\beta}
            X_{\sigma\rho}^{\prime}{\cal M}^{\sigma\rho\beta}
                     (g_1^{\prime}g_2^{\prime} \ra \overline q_4q_3g_5)^*.
                                      &\hspace{.6in} (14)
\end{array}
$$ Details will be presented elsewhere.\cite{mctobe} We focus here on the
infrared singularities, which provide a consistency check at one loop
order that ${\cal M}(G_0 \ra \qbq)$ vanishes in the chiral limit.

In general there could be soft IR divergences for $E_q, E_{\overline
q}, E_g \ra 0$ and collinear divergences for $\theta_{qg},
\theta_{\overline qg}, \theta_{\qbq} \ra 0$. In fact, only the $\qbq$
collinear divergence occurs, as can be seen from the distributions in
figure 1, obtained by imposing only the cut $\theta_{\qbq} > 0.1$ in
the $G_0$ rest frame: neither $dN/dE_g$ nor $dN/dE_q$ diverge at low
energy, and only $dN/\theta_{\qbq}$ diverges at $\theta_{\overline qq}
\ra 0$. Instead $dN/dE_g$ diverges at the {\em maximum} energy, $E_g =
m_{G_0}/2$, and $dN/d\theta_{qg}$ diverges for $\theta_{qg} \ra
\pi$. Both of these divergences are kinematical reflections of the
collinear singularity at $\theta_{\qbq} \ra 0$, for which the $\qbq$
pair with $m_{\qbq}=0$ recoils with half of the available energy
against the gluon in the opposite hemisphere.

\begin{figure}[t]
\centerline{\rotatebox{90}{\includegraphics[width= .37\textwidth]{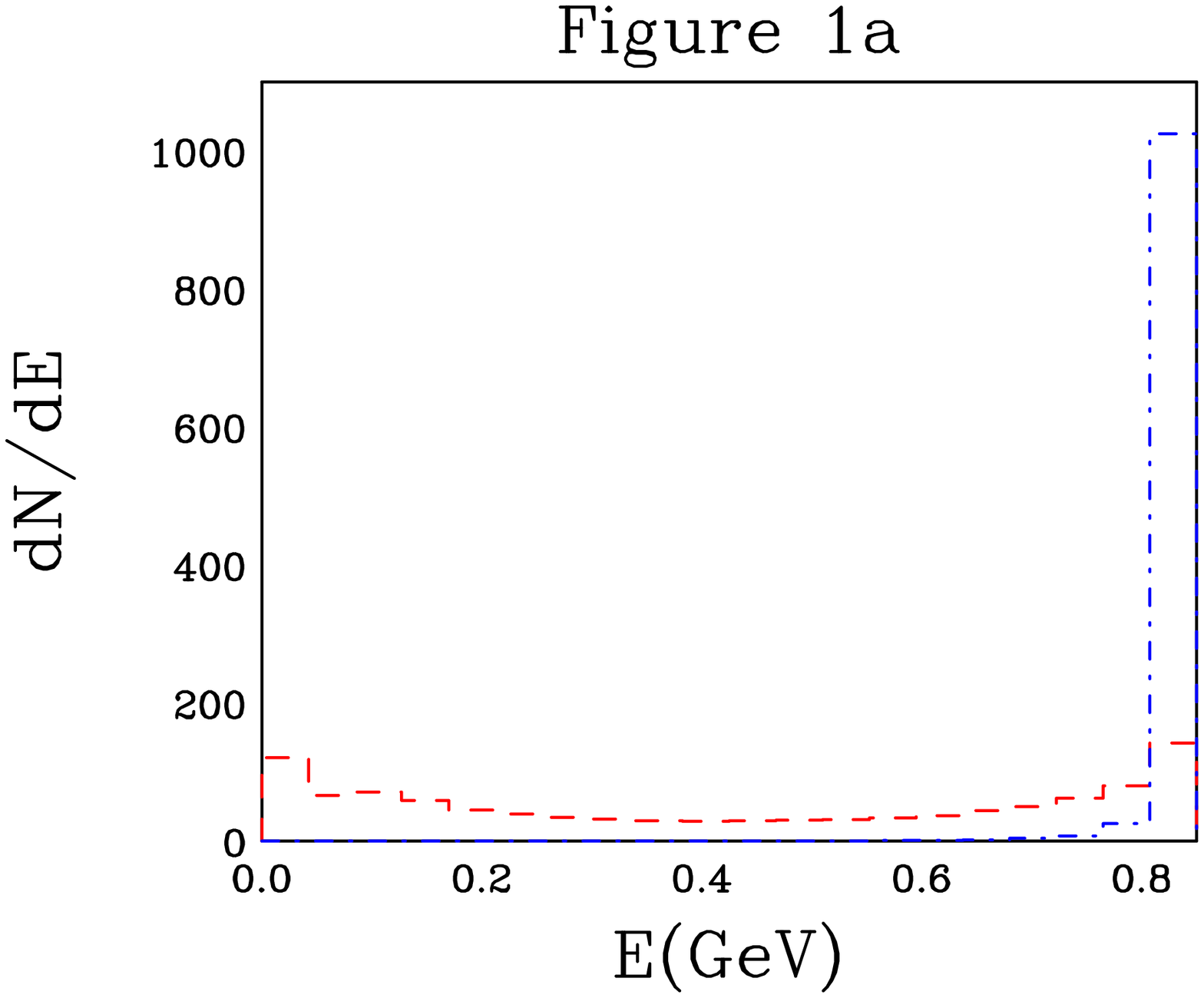}}
\hspace*{1cm}
\rotatebox{90}{\includegraphics[width= .37\textwidth]{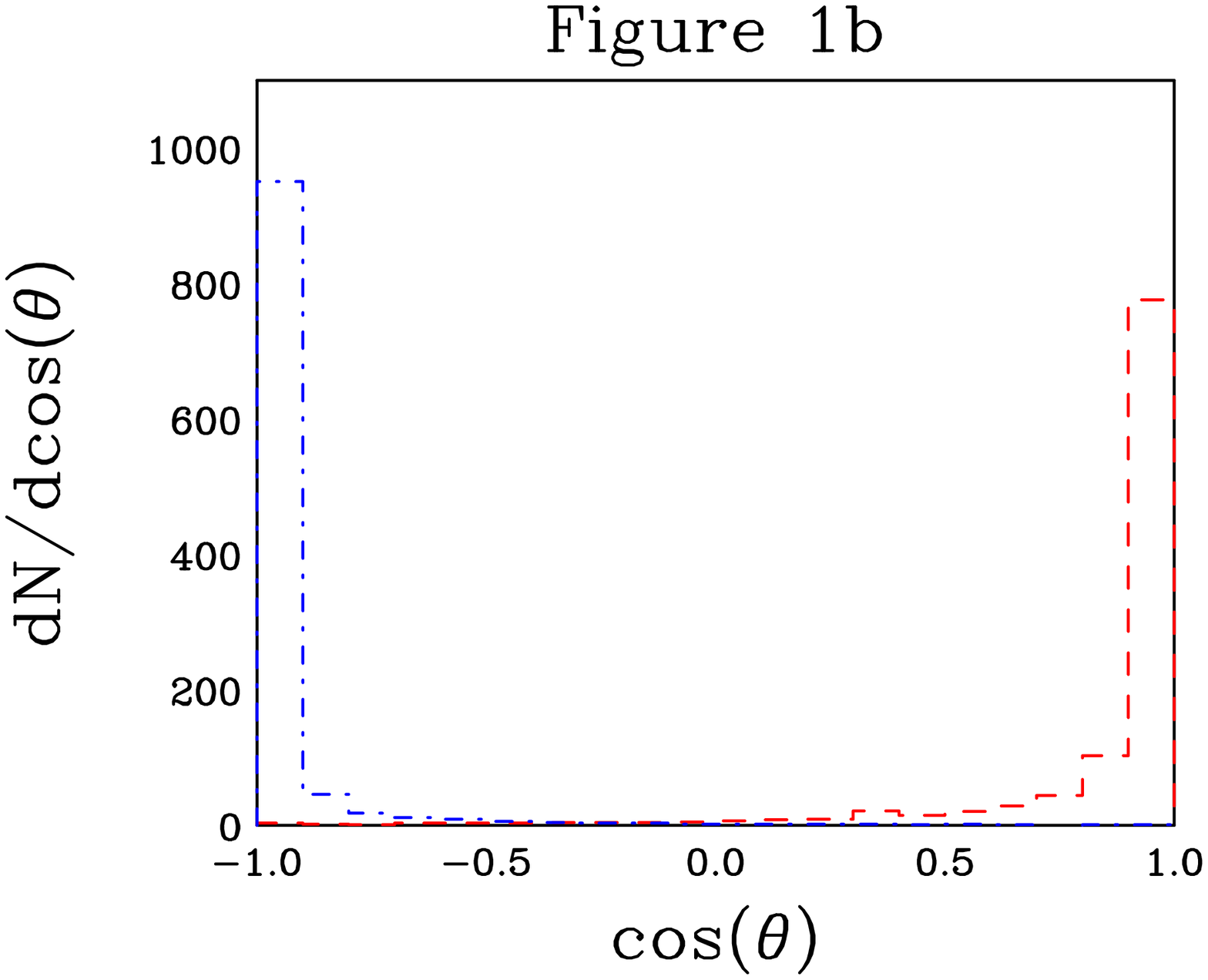}}}
\caption{Distributions for $G_0\ra \qbqg$ in arbitrary units. In figure (1a)
the dot-dashed line is $dN/dE_g$ and the dashed line is $dN/dE_q$. In 
figure (1b)
the dot-dashed line is $dN/{\rm cos}\theta_{qg}$ and the dashed line is 
$dN/d{\rm cos}\theta_{\qbq}$.}
\end{figure}

This is precisely the pattern of divergences required if $G_0 \ra
\qbq$ is chirally suppressed to all orders and, in particular, at one
loop. For if there were soft divergences in any of $E_q, E_{\overline
q},E_g$ or collinear divergences in $\theta_{qg}$ and $\theta_{\overline
qg}$, then the resulting singularities at $m_{qg}, m_{\overline qg}
\ra 0$ would have to be cancelled by virtual corrections to $G_0 \ra
\qbq$, such as gluon self energy contributions to the quark
propagator. The absence of these singularities is a consistency
check (i.e., a necessary condition) that $G_0 \ra \qbq$ is chirally
suppressed at one loop order. The collinear divergence for 
$\theta_{\qbq} \ra 0$ is cancelled by quark loop contributions to the
$gg \ra gg$ amplitude, which in the present context are one loop
corrections to the $G_0$ wave function.

{\it 5. Discussion. --- }
We have shown to all orders in perturbation theory and with a simple,
nonperturbative physical argument that the ground state $J=0$ glueball
has a chirally suppressed coupling to light quarks, ${\cal M}(G_0 \ra
\qbq) \propto m_q$, with corrections of higher order in $m_q/m_G$.
From equation (7) with $m_u,m_d,m_s$ varied within $1\sigma$
limits,\cite{mq} $\Gamma(G_0\ra \sbs)$ dominates $\Gamma(G_0\ra \ubu +
\dbd)$ by a factor between 20 and 100.  Flavor symmetry is reinstated
for $G_0 \ra \qbqg$ when the gluon is well separated from the $q$ and
$\qb$. For sufficiently heavy $m_G$ one can test this picture by
measuring strangeness yield as a function of thrust or sphericity,
with enhanced strangeness in high thrust or low sphericity events, but
it is unclear if this is feasible for $m_G \simeq 1.7$ GeV. It is 
more feasible for the ground state pseudoscalar glueball, which is 
expected to be heavier than the scalar and which we also expect to be 
subject to chiral suppression. 

For light scalar glueballs, the best hope to see strangeness
enhancement is the two body decays, $G_0\ra K^+ K^-/\pi^+
\pi^-$. 
Since $G_0 \ra \qbqg$ and $G_0 \ra ggg$ are not chirally
suppressed, naive power counting suggests $\Gamma(G_0 \ra \qbqg + ggg)
\geq \Gamma(G_0 \ra \sbs)$, so that $n\geq 3$ parton decay amplitudes
are probably the dominant mechanism for multihadron production. 
Then
$\kbk$ will dominate two body decays while multiparticle final states
are approximately $SU(3)_{\rm Flavor}$ symmetric, up to phase space
corrections favoring nonstrange final states.

Chiral suppression has a major impact on the experimental search
for the ground state scalar glueball.  Candidates cannot be ruled out
because they decay preferentially to strange final states, especially
$\kbk$, and mixing with $\sbs$ mesons may be enhanced.  This picture
of a chirally suppressed $G_0$ fits nicely with the known properties
of the $f_0(1710)$ meson. It is copiously produced in radiative $\psi$
decay in the $\psi\ra\gamma \kbk$ channel\cite{besa} and in the
gluon-rich central rapidity region in $pp$ scattering,\cite{pp} has a
small $\gamma \gamma$ coupling,\cite{gammagamma} has a mass
consistent with the prediction of quenched lattice
QCD,\cite{morningstar} and has a strong preference to decay to
$\kbk$, with $B(\pi\pi)/B(\kbk) < 0.11$ at 95\% CL.\cite{besb} As a
rough estimate of the stickiness,\cite{stickiness} we combine the
$\gamma \gamma$ 95\% CL upper limit with central values for $\psi$
radiative decay\cite{besa}, with the result $
S(f_0(1710)):S(f_2^\prime(1525)):S(f_2(1270))\simeq (> 36):12:1
$.
A more complete discussion of the experimental situation will 
be given elsewhere\cite{mctobe} --- see also \cite{beijing}.

The interpretation of $f_0(1710)$ as the chirally suppressed 
scalar glueball can be tested both theoretically and
experimentally.  Lattice QCD can test the prediction that $G_0\ra
\kbk$ is enhanced for the ground state $J=0$ glueball but not for
$J>0$. With an order of magnitude more $J/\psi$ decays than BES II,
experiments at BES III and CESR-C will extend partial wave analysis to
rarer two body decays and to multiparticle decays. They could
confirm $B(f_0 \ra \kbk)/B(f_0 \ra \pp) \gg 1$ and, if, as is likely,
the rate for multiparticle decays is big, the lower bound on $B(\psi
\ra \gamma f_0(1710)$ will increase beyond its already appreciable value
from $\kbk$ alone.  A large inclusive rate $B(\psi \ra \gamma f_0)$, a
large ratio $B(f_0 \ra \kbk)/B(f_0 \ra \pp)$, and approximately flavor
symmetric couplings to multiparticle final states would support the
identification of $f_0(1710)$ as the chirally suppressed, ground state
scalar glueball.

\noindent{\it Note added:} Cornwall and Soni\cite{cs} observed that
the trace anomaly\cite{ta} implies chiral suppression of the $G_0$
coupling to $\pp$ and $\kbk$ at {\em zero four-momentum}; however, the
extrapolation to the $G_0$ mass-shell, e.g., at $\simeq 1700$ MeV, is
not under control and could be large. Similarly, in the conjectured
AdS/CFT approach to QCD the scalar glueball is a dilaton,\cite{ads}
which therefore has chirally suppressed two body couplings at zero
momentum. I thank A. Soni, H. Murayama, and M. Schwartz for bringing
this work to my attention.
\vskip 0.2in
\noindent {\bf Acknowledgements}: I wish to thank M. Golden and S. Sharpe
for helpful discussions.


\noindent{\small This work was supported in part by the Director, Office of
Science, Office of High Energy and Nuclear Physics, Division of High
Energy Physics, of the U.S. Department of Energy under Contract
DE-AC03-76SF00098}



\end{document}